Notice: This manuscript has been authored by UT-Battelle, LLC, under Contract No. DE-AC0500OR22725 with the U.S. Department of Energy. The United States Government retains and the publisher, by accepting the article for publication, acknowledges that the United States Government retains a non-exclusive, paid-up, irrevocable, world-wide license to publish or reproduce the published form of this manuscript, or allow others to do so, for the United States Government purposes. The Department of Energy will provide public access to these results of federally sponsored research in accordance with the DOE Public Access Plan (http://energy.gov/downloads/doe-public-access-plan).

# Learning from imperfections: constructing phase diagrams from atomic imaging of fluctuations

Lukas Vlcek[1,3] Maxim A. Ziatdinov[2,3], Alexander Tselev[4], Arthur P. Baddorf[2,3],

Sergei V. Kalinin[2,3*], Rama K. Vasudevan[2,3**]

[1]Chemical Sciences Division, [2]Center for Nanophase Materials Sciences, and [3]Institute for Functional Imaging of Materials, Oak Ridge National Laboratory, Oak Ridge TN 37831, USA

[4]Department of Physics & CICECO- Aveiro Institute of Materials, 3810-193 Aveiro, Portugal

**Abstract**

Materials characterization and property measurements are a cornerstone of material science, providing feedback from synthesis to applications. Traditionally, a single sample is used to derive information on a single point in composition space, and imperfections, impurities and stochastic details of material structure are deemed irrelevant or complicating factors in analysis. Here we demonstrate that atomic-scale studies of a single nominal composition can provide information on a finite area of chemical space. This information can be used to reconstruct the material properties in a finite composition and temperature range. We develop a statistical physics-based framework that incorporates chemical and structural data to infer effective atomic interactions driving segregation in a $La_{5/8}Ca_{3/8}MnO_3$ thin-film. A variational autoencoder is used to determine anomalous behaviors in the composition phase diagram. This study provides a framework for creating generative models from diverse data and provides direct insight into the driving forces for cation segregation in manganites.

*sergei2@ornl.gov

**vasudevanrk@ornl.gov

Accelerating materials discovery is critical to solving many of the world's challenges as can be reflected by large-scale endeavors such as the materials genome initiative.[1-3] The traditional materials discovery workflow relies on individual sample preparation and characterization, effectively resulting in a sampling of a single point in a chemical space per sample synthesized. The heterogeneities such as chemical segregation, point and extended defects, impurity phases and the like are traditionally perceived as a limitation that negatively affects the veracity of the data, and synthesis strategies are traditionally aimed at avoiding these.[4] The inherent atomic disorder such as stochastic atomic configurations are perceived to be self-averaging and irrelevant to macroscopic properties. However, for materials such as ferroelectric relaxors, charge-ordered manganites, or filamentary superconductors, nanometer scale heterogeneities are defining features critical to understanding the physical functionality, and still remain one of the hot topics in condensed matter physics.[5,6] [7]

This research paradigm results in large inefficiencies as the process is time-consuming, and the method of sampling is largely driven by intuition. In the modern era, there have been many attempts to alter this paradigm to accelerate materials discovery: one method is to replace synthesis of an individual composition with a range of compositions, through a high-throughput combinatorial approach, and which has seen substantial successes in the recent past.[8-11] Other approaches rely on use of theory-guided materials discovery, wherein first-principles calculations are performed to narrow down the potential list of candidate materials to investigate.[12,13] The latter method explains the rise of libraries such as the Materials Project, AFLOWLIB[14], JARVIS-DFT[15], Polymer Genome[16], and many others. These calculations can be supplemented with results from experiments within a Bayesian framework for experimental design optimization[17-19], to reduce the number of experiments taken to arrive at the optimal material given a set of input (processing) parameters.

Here we pose that by studying the characteristic structural and chemical fluctuations that exist within a single chemical composition, we can infer the relevant interactions and produce a generative model that can predict properties in a finite region of the chemical and temperature space. This approach is rooted in the principles of statistical physics, in that any macroscopic system can be understood in relation to the interactions between its constituent components, and understanding these interactions enables predictions of the system as a function of thermodynamic state variables, including temperature, pressure, and chemical potential. The key insights can be gained via analysis of the fluctuations of the system, which has commonly been used in cases such as biomolecule unfolding via atomic force microscopy[20,21], or more recently, studying the dynamic trajectories of DNA molecules in a nanofluidic environment, essentially turning measurements of stochastic trajectories into equilibrium free energies.[22] In our work, we aim to use the compositional and structural fluctuations in the quenched (static) system to build a generative model encoding the effective interactions in the system.

Although feasible in principle, in practice the determination of the form and values for the interactions within generative models to match experiment is a difficult task. For example, if one considers a 2D Ising system on a square lattice, the model contains only a small number of input parameters and can be sampled to produce configurations (**Figure 1**(a)). Bulk property measurements on many samples (indicated by the blue circles) can be carried out across the temperature-composition space, and the parameters of the generative model can be adjusted based on experiment, to infer the likely configurations within individual samples. On the other hand, if imaging modalities are available, the alternative approach as shown in Fig. 1(b), becomes practical.

In this case the imaging studies can constrain the generative model's parameters to match the observed configurations with those generated from the model, and then the generative model can be used to predict the configurations as a function of e.g. temperature and composition. The use of machine learning (specifically, variational autoencoders) on these generated configurations provides an effective tool to map any anomalies expected in the phase diagram, even though only one material of a specific composition was characterized. Of course, one expects that the uncertainty in these phase diagrams to increase as the composition or temperature departs further from the actual measurement, but this principle holds so long as the generative model is adequate to represent the interactions present in the system.

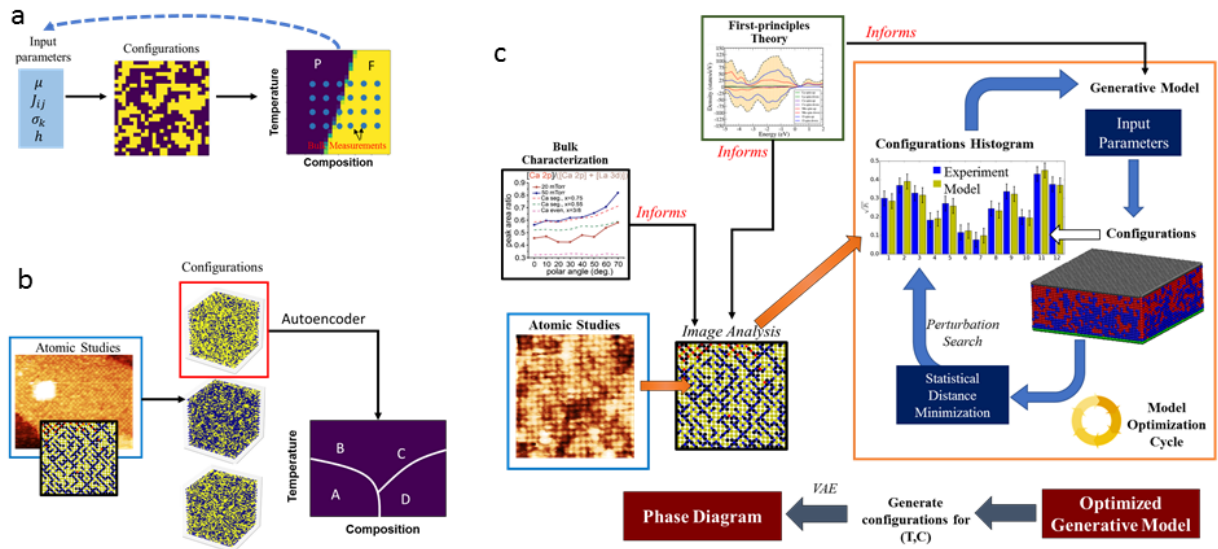

**Figure 1: Phase diagram determination workflows.** **(a)** Generative models exist, but are difficult to fit to macroscopic data, and require many compositions and temperature spaces to be explored. **(b)** An alternative is available if imaging data is present, because the statistics of the configurations found in experiment form the imaging can be compared with those found in the generative model and used for inference of model parameters. **(c)** Our proposed workflow involves incorporating data from atomic-scale imaging with macroscopic measurements and first-principles theory, in a robust framework underpinned by statistical mechanics. Atomic resolution form STM or STEM provides information on location and composition of the atoms. Classification can be improved via insight from theory, and further by constraining the composition of the surface layer. Subsequently, a generative model is optimized to minimize the statistical distance between the model and the experimental data. Again, macroscopic measurements constrain the range of the generative model. The result after optimization is a model that can be used to predict the system behavior as a function of e.g. chemical potential or temperature, and these data can be fed directly into an autoencoder to determine any anomalies in the phase diagram, to highlight points for further investigation

Here, we demonstrate this approach for a model system, focusing on elemental segregation in a $La_{1-x}Ca_xMnO_3$ thin film, a material which is relevant for oxide fuel cell applications and electronics, and which has been studied previously by both experimental[23-25] and combined experimental-theoretical[26] approaches. We show that experimental data from scanning tunneling microscopy images can be consistently combined with electron spectroscopy data (providing a chemical profile as a function of depth) on thin-film $La_{5/8}Ca_{3/8}MnO_3$ to infer the parameters of a 3D generative statistical mechanical model that provides insights into sub-surface microstructure and the material's thermodynamic response. The generative model is used to predict the configurations expected within the surface and bulk across the composition and temperature space and is subsequently analyzed via a variational autoencoder approach to determine anomalies within this space. We find that the model of effective metal-metal interactions incorporating the two independent sources of experimental information leads to the prediction of weak segregation forces in bulk (attributable to elastic effects) and weak de-segregation forces in the surface (attributable to electrostatic interactions), in agreement with quantum-chemical calculations.[26] In addition, we quantify the influence of surface oxygen on elemental segregation. This study marks a significant advance in the materials discovery paradigm, leveraging recent major advances in imaging, machine learning and statistical physics, and will be useful in accelerating materials optimization, targeted design and materials discovery.

**Results**

The outline of the method used in this letter is shown in Figure 1(c). Atomic-scale images in conjunction with macroscopic data, e.g. from x-ray photoelectron spectroscopy, are measured on the sample of interest. If first principles calculations are available, they can be utilized in the analysis stage of the images, for e.g. aiding in atom classifications.[27] Chemical composition can also assist in providing constraints on the image analysis, to ensure that the macroscopic averages are maintained. These constraints to reduce error during classification of inherently noisy electron or scanning tunneling microscopy images, albeit it should be noted that certain care is required in applying them (e.g. surface segregation can affect surface chemical composition compared ot average, affecting STM but not STEM images).

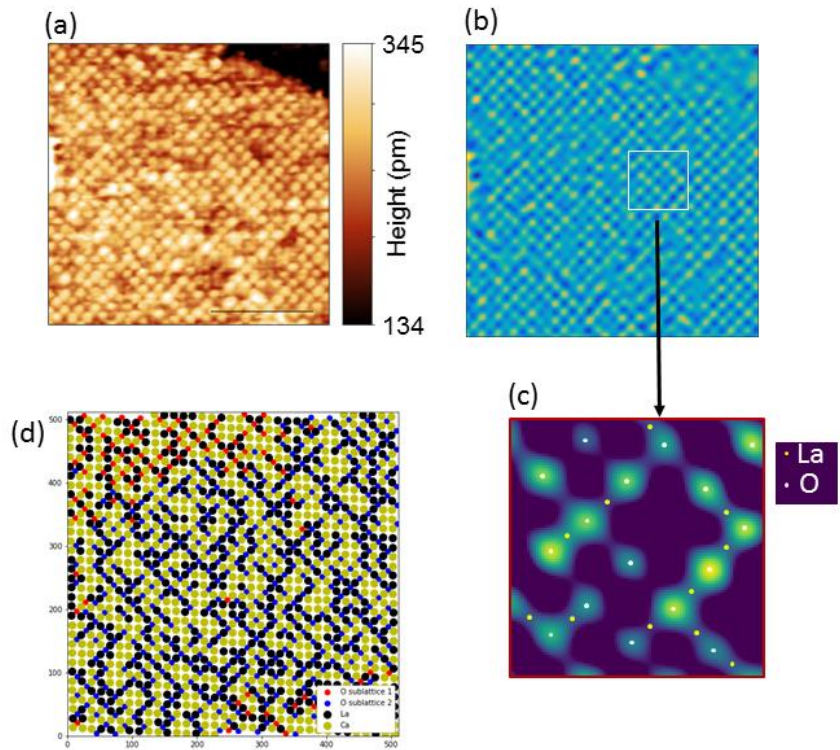

**Figure 2: In-situ STM image and analysis of the A-site terminated surface of La$_{5/8}$Ca$_{3/8}$MnO$_3$ thin film**. (a) STM image (scale, 5nm) of the A-site terminated surface showing a ($\sqrt{2}$ x $\sqrt{2}$) R45° reconstruction. (b) Image after filtering with PCA sliding window denoising algorithm. (c) Expanded view of region in (b), with simple thresholding to better locate the atom centers and lobes of intensity. The bright atoms are the Oxygens, and lobes of intensity around each O can sometimes be seen. If the intensity within these lobes is above a certain threshold, a La atom is placed; else, a Ca atom is presumed present. (d) Fully analyzed image showing the O sublattice, along with the La and Ca.

After the positions and types of each atom is identified, histograms of the configurations are computed. A generative model of a certain complexity of interactions is setup and optimized via a perturbation search over the model parameter space to minimize the statistical distance of the histograms between the target system and the model. The optimum model then provides the inferred model parameters for interaction energies and can be used to predict the system behavior under different thermodynamic variables. This knowledge can further be used in an unsupervised manner to build the phase diagram of the system. For example, analysis of the generated configurations can be achieved via the use of variational autoencoders, which have been shown to be useful in delineating phase transitions in both static and dynamic models in the recent past.[28,29] By quantifying the elementary driving forces within a generative model, we can subsequently predict the system's response to changing thermodynamic conditions, such as temperature or chemical potential, and optimize these conditions to produce microstructures associated with desired materials properties.

**Elemental segregation in thin-film perovskite from microscopic imaging and spectroscopy**

Information about the distribution of La and Ca atoms in the surface layer of perovskite $La_{5/8}Ca_{3/8}MnO_3$ thin film was extracted from *in-situ* STM images, informed by density functional theory (DFT) calculations (unpublished). In the present case the film was grown by pulsed laser deposition on (001) etched $SrTiO_3$ substrates and is of thickness ~50 unit cells (see Methods). We have previously explored the growth, surface reconstructions and terminations of these films through a combination of *in-situ* atomic-scale imaging, low-energy electron diffraction, and angle-resolved x-ray photoelectron spectroscopy (XPS), as published elsewhere.[30,31] As shown in Fig. 2(a), the surface of the (La,Ca)-terminated film is reconstructed into a ($\sqrt{2}$ x $\sqrt{2}$)R45° lattice, with every second unit cell missing an apical oxygen atom. Because these oxygens contain the highest density of states at the Fermi Level, they are highly visible in STM images.[30] The raw images were cleaned using the principal component analysis-based filtering, and the position of the Oxygen atoms were found using a motif-matching approach.[32] This algorithm is freely available through the open-source pycroscopy package.[33]

Interestingly, both in raw and cleaned images a fine structure can be seen around some of the oxygen atoms, corresponding to the electronic density (intensity) somewhat elongated in the directions of presumed metal (La, Ca) atoms, as seen in Fig. 2(b,c) (these are also visible in raw images, see Fig S1). This observation is consistent with results of our DFT calculations (unpublished) of charge density (as well as with basic chemical considerations),[34] which predicted a higher electron density located between oxygen and $La^{3+}$ as compared to $Ca^{2+}$. Therefore, we used the DFT-based insights to associate these lobes of higher STM intensity with the presence of neighboring $La^{3+}$ atoms. This information along with the known relative La/Ca concentration determined from XPS experiments allowed us to assign the positions of La and Ca in the surface layer, as shown in Fig. 2(d), with O atoms shown in blue or red (depending on the sublattice), La atoms in black and Ca atoms in yellow. In this way we constrained the analysis based off both first-principles calculations as well as bulk characterization. Interestingly, many of the La and O atoms form structures resembling self-avoiding paths or a maze, rather than forming segregated clusters or being completely randomly distributed, which can be seen in the contours in Fig. 2(d). Note that the missing regions of the upper right and lower left quadrants are deliberately not analyzed further due to large concentration of defects in these regions. These patterns indicate specific interactions between the surface species. To capture this behavior for the use in subsequent model optimization, we collected the histograms of local configurations, while distinguishing the configurations according to the presence or absence of a surface oxygen atom. The histograms are shown in Fig. 3(a,b) for the configurations with and without the surface oxygens, respectively and plotted as yellow bars. In these plots the specific configurations are plotted above the histogram.

The higher frequency of configurations 5 and 6 in the presence of oxygen in Fig. 3(a) relative to Fig. 3(b) indicate stronger affinity of O atoms to La compared to Ca. Also, the increased frequency of configuration 4 in the presence of oxygen indicates that La atoms tend to align on the opposite sites of the oxygen, minimizing their electrostatic repulsion. Note that the histograms contain error bars (1 standard deviation) obtained by undertaking the same image analysis for three independent images shown in Supplementary Figures 2 -4.

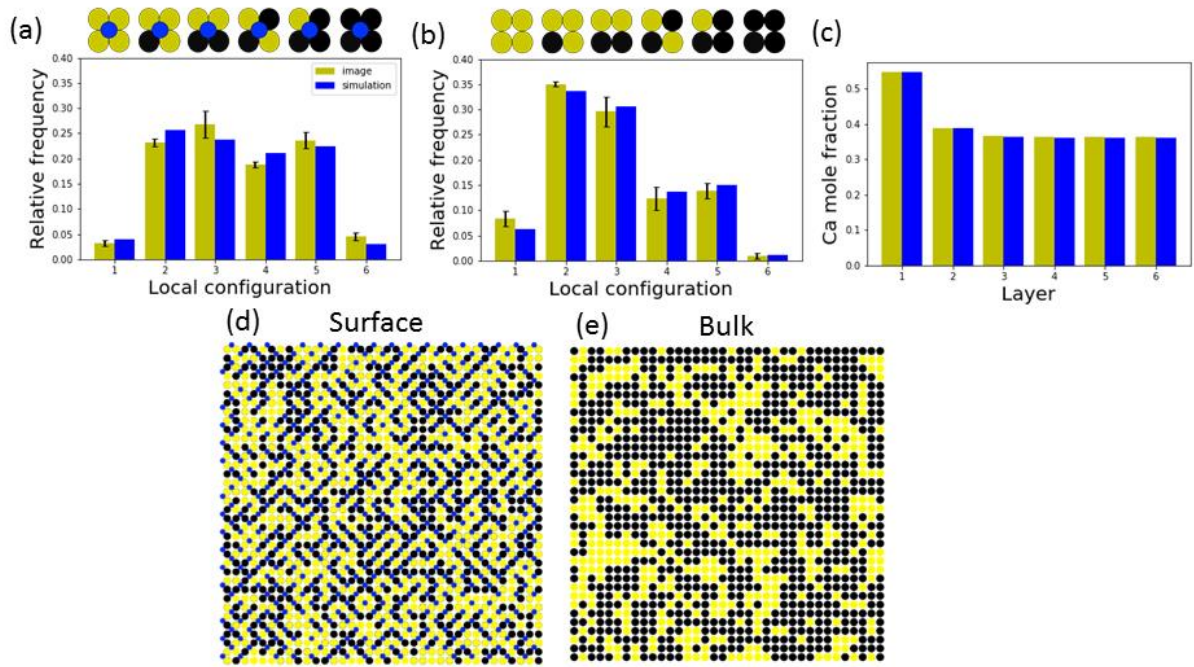

**Figure 3: Structural statistics for LCMO thin film from experiments (yellow) and simulations (blue).** Histograms of local configurations of metal atoms surrounding a surface oxygen **(a)** and configurations without a surface oxygen **(b)**. **(c)** The concentration profile of Ca atoms across 6 top layers. The optimized generative model appears to very closely match the experimentally observed parameters, indicating it is of sufficient complexity to model the interactions. The snapshots of atomic configurations in the top (d) and bulk (e) layers from MC simulations of the LCMO thin film. $La^{3+}$ - black, $Ca^{2+}$ - yellow, surface O – blue.

While the information from STM imaging provides detailed information about the atomic distributions in the top layer, it may not provide sufficient constraints on the elemental segregation deeper within the thin film, especially if the top layer can be expected to have distinctive properties and composition due to segregation, etc. For this reason we complemented the imaging data with angle-resolved XPS measurements that allowed us to infer the concentration profile across the layers of the thin film.[30] This data shows that for this particular film, the surface concentration is ~ $La_{0.5}Ca_{0.5}MnO_3$, and at more direct angles, the bulk concentration is recovered. Quantitative analysis of angle-resolved XPS data is difficult due to large number of unknown parameters in modeling, but experiments on LCMO show that the segregation is likely limited to only two-three unit cells at the surface, as in the study by Plummer *et al*.[35] Therefore, the concentration profile was assigned as varying exponentially from $x = 0.54$ to $x = 0.375$ three unit cells into the film, corresponding roughly to the correlation length of 0.5 unit cells as found in ref 35.

**Model construction**

The statistical information collected from the experiments serves as the primary constraint on the model parameters. The form of the model was chosen to respect the known physics, as well as to generate data comparable with the experimental target and provide the predictions of interest. With the goal of predicting atomic-scale structures and the response to changes in thermodynamic conditions, it is necessary to go beyond mean-field descriptions and construct a model capable of capturing fluctuation properties. Also, the fact that the XPS data show clear segregation between bulk and the top layer, which constantly changes as the thin film grows, indicates that the metal atoms are mobile and able to adjust to the different thermodynamic gradients in bulk and surface. Therefore, we will assume a statistical mechanical model of an equilibrium system. Given the considerations of experimental and statistical uncertainty in the target data, our guiding principles are the model's physical plausibility and its simplicity that would prevent overfitting. In the construction of the model form, we can build on prior insights made by others as well as on our own studies. Lee and coworkers identified and analyzed two main contributions to elemental segregation in manganate perovskites, elastic and electrostatic, and argued that the former will vanish within the surface layer,[26] while other studies have detected metallic character of the top layer[34] for a particular LCMO surface,[36] which may uniquely influence the interactions among the top-layer atoms. To accommodate these features while keeping the model simple, we chose to represent the thin-layer system by a lattice model with effective pairwise interactions that can control metal atom segregation across bulk and surface layers and take into account the presence of surface apical oxygens. Since here we are only interested in the segregation of $Ca^{2+}$ and $La^{3+}$ ions, we can limit the translational mobility to these particles, while treating the Mn and O atoms as static scaffolding. The energy $u_i$ of configuration $i$ of the thin-layer system is then given as

$$u_i = w_{CL} \sum_{\{C,L\}} \sigma_i^{b,CL} + w_{LL} \sum_{\{L,L\}} \sigma_i^{s,LL} + w_{LO} \sum_{\{L,O\}} \sigma_i^{s,LO} \qquad (1)$$

where $C$ ($Ca^{2+}$), $L$ ($La^{3+}$), and O denote the two types of the metal cations and surface oxygens, respectively; the indicators $b$ and $s$, denote whether these cations are located in the bulk or surface layers. The quantities $\sigma_i^{XY}$ are either 1, if the particles are of types $X$ and $Y$, or 0 otherwise. Here the parameter $w_{CL}$ defines the energy of interactions between the nearest-neighbor pairs of $Ca^{2+}$ and $La^{3+}$ cations in bulk, $w_{LL}$ defines the interactions between surface $La^{3+}$ cations unshielded by surface oxygens, and $w_{LO}$ defines the interaction of $La^{3+}$ with the nearest surface oxygens. To calculate the system properties, we used canonical Monte Carlo simulations.

**Translating the statistics of structural descriptors into a generative model**

To optimize the three interaction parameters, we match model predictions with the two complementary sources of information: the statistics of local configurations collected from image analysis, and the concentration profile across layers from XPS measurement. This is performed within a newly developed statistical mechanical framework based on minimization of the statistical (or information) distance metric.[37-39] This approach presents a key advance in the design of statistical mechanical models, as it is designed to be robust against sampling uncertainty and, unlike other proposed approaches (e.g., simple least squares fitting, relative entropy minimization)[40] allows the model to properly capture the target system's fluctuations.[39,41] As follows from response theory, thermal fluctuations contain information about the response of the system to external perturbations (e.g., temperature, chemical potential), and thus their reproduction is essential for the model to be predictive. The statistical distance loss function for the set of target histograms is defined as,

$$S^2 = c_{SO}s_{SO}^2 + c_{SN}s_{SN}^2 + c_{PR}s_{PR}^2 \quad ; \quad s = \arccos\left(\sum_{i=1}^{k}\sqrt{p_i}\sqrt{q_i}\right) \quad (2)$$

where the subscripts $SO$ and $SN$ denote surface configurations with and without oxygen, respectively, and $PR$ denotes the concentration profile; $p_i$ and $q_i$ are the probabilities of finding outcome $i$ in the measurement of the experimental system $P$ and model $Q$, respectively, with the total of $k$ possible outcomes. The specific forms of the statistical distances $s^2$ for different target data are given in Supplementary Note 1. The coefficients $c_X$, representing the relative weights of the $SO$, $SN$, and $PR$ data, were taken here as equal. Besides parameter optimization, we used the metric to select the functional form of the model that best reproduced the target data with minimum number of adjustable parameters. In case the statistical distance is not minimized (to some threshold), the model can be re-formulated to consider more interactions, otherwise the model is kept as simple as possible.

The optimal model parameters minimizing Eq. (2) were found using a computationally efficient perturbation technique,[39,42] which takes advantage of the thermodynamic perturbation relations to predict the changes of the model generated histograms of local configurations and concentration profiles without needlessly repeating the actual simulations. The landscapes of the statistical distance ($S^2$), whose minimum we seek, are shown as functions of the interaction parameters in Fig. S4. The optimization yielded effective interactions in reduced units ($w' = w/k_B T$) $w'_{LC} = 0.194$, $w'_{LL} = 0.739$, and $w'_{LO} = -0.890$ kJ/mol. Assuming that the system was equilibrated at the synthesis temperature of 1023 K, we can assign absolute values to the parameters as $w_{LC} = 1.65$ kJ/mol, $w_{LL} = 6.28$ kJ/mol, and $w_{LO} = -7.57$ kJ/mol. The resulting model can reproduce quantitatively the target distributions of local surface configurations as well as concentration profiles derived from the experimental data, which are plotted in Fig. 3(a-c) with the experimentally observed results in yellow, and the simulation results in blue. Sample configurations of the surface and bulk generated from this optimized model are shown in Fig. 3(d) and 3(e), respectively.

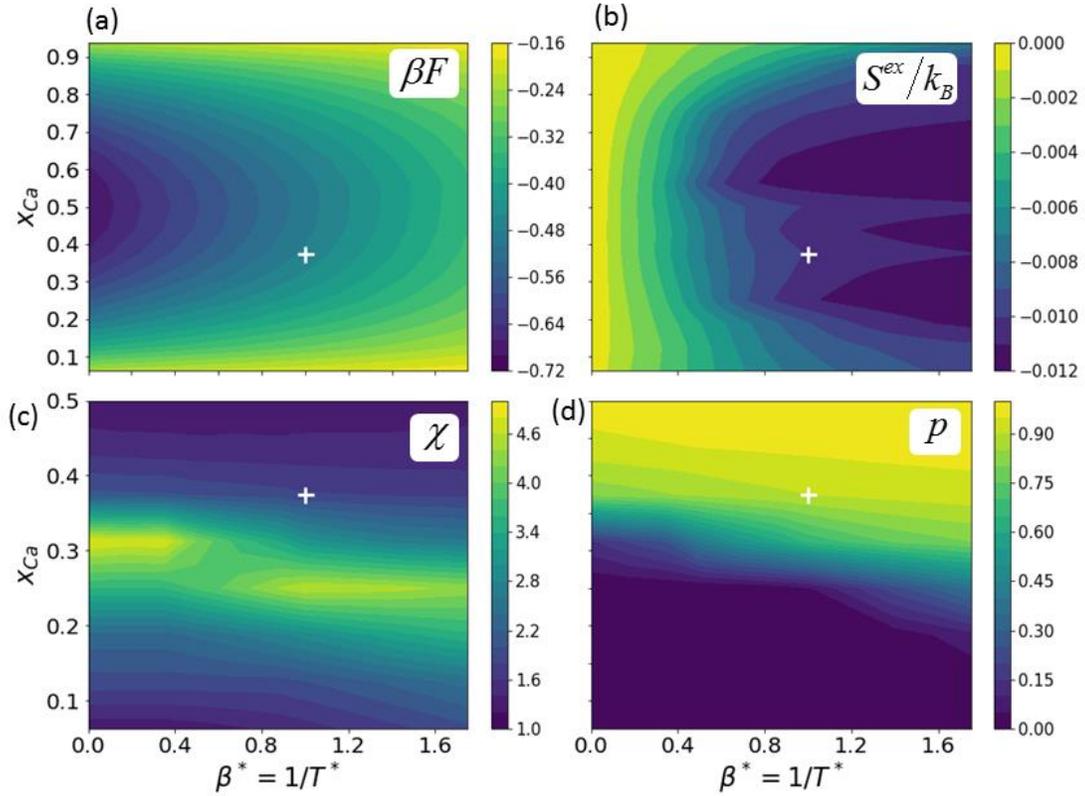

**Figure 4: Model predictions: thermodynamics and structural properties of the LCMO thin layer as a function of reduced inverse temperature and elemental composition.** (a) Reduced Helmholtz free energy of elemental segregation, $\beta F$, and (b) excess entropy of elemental segregation, $S^{ex}/k_B$. (c) The mean size of finite Ca clusters, $\chi$; (d) probability $p$ of Ca being a part of infinite cluster (spanning the periodic simulation box). White cross indicates our experimental conditions.

The resulting interaction parameters can be interpreted in terms of elastic and electrostatic effective interactions. A systematic study[26] of contributions to dopant segregation in manganate perovskites predicted that the elastic contribution to the effective interactions can be expected to be small because the size of $La^{3+}$ cations (1.172 Å) is only slightly larger than that of $Ca^{2+}$ (1.14 Å),[43] while the electrostatic effect on the surface segregation depended strongly on the considered model and ranged from positive to negative energy contributions. In agreement with this study,[26] we identified unfavorable electrostatic interactions between the like cation pairs as a major, but not the only effect, driving the structuring within the surface layer, and favorable elastic interactions, which prevail in the deeper layers and lead to the increased tendency to segregation. As an addition to these driving forces, we have identified and quantified the effect of surface oxygens on the metal atom distribution.

**Predicting subsurface segregation and the thermodynamic response**

While our simple generative model can reproduce the experimental observations, its main value is that it allows for generalization, i.e. making fast predictions of the system properties for different elemental compositions and temperatures, which can be then used for inverse design of materials and identifying most informative experiments. For instance, certain microstructures can be identified by quantum chemical methods as having desirable electronic properties, and our modeling can predict conditions at which these will be generated. Here we predict some basic thermodynamic and structural characteristics as a function of the external parameters. We used special sampling techniques (see SI) to efficiently explore the materials property space as functions of the synthesis temperature $T$ and elemental composition in terms of Ca/(Ca+La) fraction $x$. In Figs. 4(a,b), we show the calculated free energy and entropy landscapes as the function of temperature and relative concentration of Ca atoms, $x$. The decrease of excess entropy for lower temperatures (*i.e*, high reduced inverse temperature $b^*$) and intermediated compositions indicates tendency for structuring, but shows no signs of phase transitions, such as abrupt changes, within the probed temperature range. The structural characteristics of the elemental segregation in terms of cluster formation for varying temperature and composition are summarized in Figs. 4(c,d) in terms of mean size of finite clusters, $C$, and the probability $p$ of Ca particle belonging to an infinite cluster spanning the simulated system. Such predictions can be used for rational design of materials with desired level of elemental segregation, percolation, or other microstructural characteristics.

**Exploring the phase diagram in an unsupervised manner**

In addition to the insight into the driving forces of elemental segregation, we use the generative model in conjunction with machine learning methods to determine interesting regions of the composition-temperature space. For example, even in the absence of macroscopic phase transitions, there may be anomalies at certain compositions or temperatures in terms of the resulting configurations. Ordinarily, investigating this would require a collection of samples and many macroscopic measurements at different temperatures, to build such a picture. In addition, even in those cases only the macroscopic characteristics are known, rather than specifics of the underlying atomic configurations and their statistical distributions. Using the generative model, several hundred configurations were computed at distinct temperatures and concentrations. This large dataset is ideally suited towards clustering-type approaches to determine the interesting regions of the composition-temperature phase space. We turned to the use of a variational autoencoder[28], which belongs to the class of neural network-based algorithms that have been shown to be useful in data compression, image generation, and recently, learning order parameters for phase transitions.[44-46] The idea is to learn a representation of the data such that it can be expressed with a small number of latent parameters. Details of the autoencoder network are available in the Methods section. Briefly, the autoencoder consist of an 'encoder', which takes input data which is passed through layers of decreasing numbers of nodes, such that a 'bottleneck' is reached (thereby learning to encoder the system as a small number of latent parameters $z$), and a 'decoder', which is akin to an encoder in reverse that attempts to learn the mapping from latent parameters to the original data. The configurations serve as both the input and the output during the training, with the optimization of the parameters minimizing the reconstruction loss. It should be noted that *variational* autoencoders are generative models, because in this case instead of learning individual latent parameters, they are reparametrized as Gaussian distributions and minimized during training with respect to both mean-squared error of the prediction from the target, as well as with the Kullback–Leibler divergence from a unit Gaussian.[47,48] After training, sampling from the distribution of $z$ enables generation of the configurations.

The results of the autoencoder output are shown in Figure 5(a), with the plot of the latent parameter as a function of temperature and effective temperature (colored). Although no clear trends can be seen along the temperature axis, a kink appears at a concentration value of x=0.50, suggesting that this area of the phase space presents an anomaly. Interestingly, the latent variable appears to map to the free energy diagram in Figure 4(a). To examine this in detail, the configurations for concentrations of $x = 0.375$ and $x = 0.50$ as predicted from the model are shown in Fig. 5(b,c), respectively. The surface structure and a slice from the bulk, 8 unit cells into the film, are shown for the two concentrations in Fig. 5(d,e). It is clear that the behavior is different for the c=0.5 concentration, with substantially less tendency for segregation in the bulk, and in agreement with the predictions made in Figure 4. Therefore, the variational autoencoder can be used as a suitable proxy to determine the areas of interest in the phase diagram – one which results from a single measurement in a chemical space, but where interactions are understood from the physics-appropriate integration of available data with generative models.

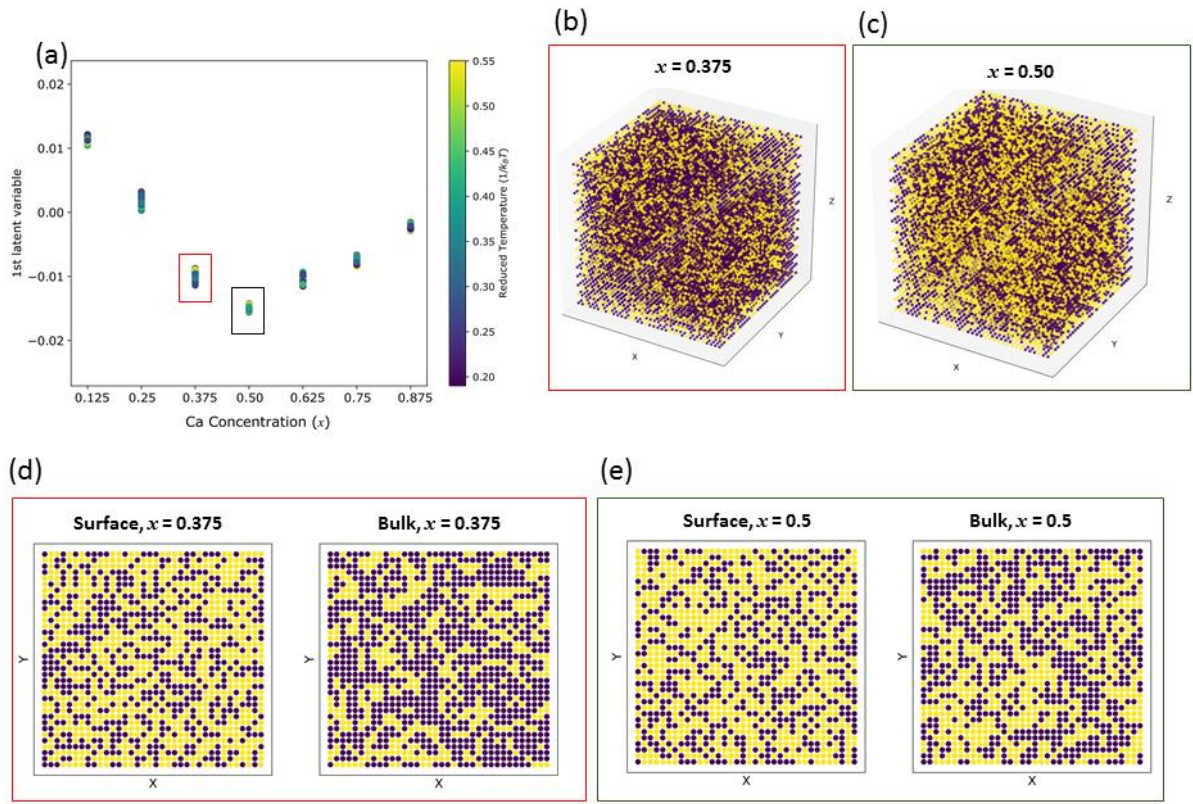

**Figure 5: Variational autoencoder for phase diagram prediction.** (a) Latent parameter as a function of composition, for different temperatures (colored). A minimum at the x=0.5 concentration is found. A configuration sampled from the x = 0.375 and x = 0.50 data are shown in (b) and (c). (d) and (e) plot the surface and bulk *x-y* slices from these configurations (purple = La, yellow = Ca).

**Discussion**

This study has substantial implications for manganites as used in solid oxide fuel cell cathodes, as well as for their electronic applications such as in tunnel junctions[49], given the strong sensitivity of the oxygen reduction reaction (and electronic properties) on the composition of the manganite surface.[50-54] The ability to understand the driving forces for segregation in both bulk and surface via a simplified model is a powerful tool, especially in systems that are often difficult to predict with first principles methods (such as those studied here).

The model development approach allowing us to predict materials properties from microscopic images and other supplementary experiments is not specific manganites and can easily be applied to other more complex cases, especially as it does not in principle require atomic-level imaging. That is, it is equally possible to consider coarse-grained collections of atoms as larger individual units for which structural statistics are computed and would be needed to understand e.g. polycrystalline materials. Additional model constraints can be devised either from first-principles simulations (e.g., in the analysis of the images step[55]) to better refine the histogram generation, or from additional characterization, such as various defect concentrations as a function of layer. On a more general level, the presented methodology for constraining model parameters by multidimensional (and multimodal) spectral data can be readily adapted to many problems in materials science. The structural modeling and selection of promising material candidates can be directly followed by quantum chemical calculations to determine the electronic properties. Undoubtedly the key to this is the development of workflows to merge experimental data with the appropriate level of theory, as has been discussed by many groups recently.[56-58] Further work could involve ensemble averaging, for more accurate uncertainty quantification, as well extension to dynamic data and associated dynamic generative models such as kinetic Monte-Carlo based methods.

**Acknowledgements**

The work was supported by the U.S. Department of Energy, Office of Science, Materials Sciences and Engineering Division (R. K. V., S. V. K., L.V). Research was conducted at the Center for Nanophase Materials Sciences, which also provided support (A.P.B.) and is a DOE Office of Science User Facility.

**Methods**

Methods, including statements of data availability and any associated accession codes and references, are available in the online version of this paper. Full details of sample preparation STM imaging, and XPS measurements can be found in our previous work.[30] Brielfy, the LCMO films were grown by pulsed laser deposition with a KrF excimer laser of fluence ~2J/cm$^2$ striking a target of nominal composition La$_{5/8}$Ca$_{3/8}$MnO$_3$, on TiO$_2$ terminated (001) SrTiO$_3$. The films were deposited at 750C at 20mTorr O$_2$ background pressure. STM imaging was performed in-itu at room temperature in an Omicron VT STM system operating at a pressure <3x10^-10 Torr. XPS Measurements were performed in-situ using nonmonochromatic Al K$\alpha$ radiation ($h\nu = 1486.6$ eV) from a SPECS XR50 Mg/Al double-anode X-ray source operated at 280 W using a SPECS PHOIBOS 150 hemispherical electron energy analyzer at room temperature in UHV at a base pressure $<8 \times 10^{-11}$ Torr.

**Simulation**

The trajectory of configurations of the model system was generated using canonical Monte Carlo (MC) simulations carried out in a rectangular box with two periodic dimensions $L_X = L_Y = 36$ and non-periodic dimension $L_Z = 16$ at reduced temperature $k_B T/\varepsilon = 1$, where $\varepsilon$ is the energy unit scaling interaction parameters. The total number of $n=10^6$ independent samples was collected from each simulation consisting of $10^8$ individual MC steps, where each step consisted of swapping the chemical identities of randomly chosen nearest neighbor particles. The simulations were used to calculate the statistics of 12 distinct local configurations of metal cations in the top layer (Fig. 3), and the relative cation concentrations across thin-film layers.

The parameter optimization simulations followed the same protocol were performed in a smaller box with dimensions $L_X = L_Y = 8$ and $L_Z = 16$ over $7 \times 10^8$ MC steps.

**Autoencoder**

The variational autoencoder consisted of a three-layer encoder section containing of 256, 128 and 64 nodes in each layer, respectively. A single latent parameter was utilized, and the decoder size was the same as the encoder in reverse. Rectified Linear units (ReLU) was used as activation functions for all layers, and dropout[59] was used on the first and last layer to avoid overfitting. We used the Adam optimizer and implemented the autoencoder using keras[60] with Tensorflow backend. Default values of the Adam optimizer were used for fitting, which was carried out for 50 epochs.

**Conflict of Interest:** The authors declare no conflicts of interest

**Supplementary Material:** Supplementary information accompanies this manuscript and contains information on the modeling, as well as more images that were analyzed.

**Author Contributions:** RKV, APB and AT performed the experiments. LV performed the statistical distance modeling, analyzed the data and co-wrote the paper. RKV, SVK and LV analyzed data and co-wrote the paper. MZ assisted with the variational autoencoder. All authors commented on the manuscript.

# Supplementary Information

**Learning from imperfections: constructing phase diagrams from atomic imaging of fluctuations**

Lukas Vlcek[1,3] Maxim A. Ziatdinov[2,3], Alexander Tselev[4], Arthur P. Baddorf[2,3],

Sergei V. Kalinin[2,3*], Rama K. Vasudevan[2,3**]

[1]Chemical Sciences Division, [2]Center for Nanophase Materials Sciences, and [3]Institute for Functional Imaging of Materials, Oak Ridge National Laboratory, Oak Ridge TN 37831, USA

[4]Department of Physics & CICECO- Aveiro Institute of Materials, 3810-193 Aveiro, Portugal

1. **Image analysis**

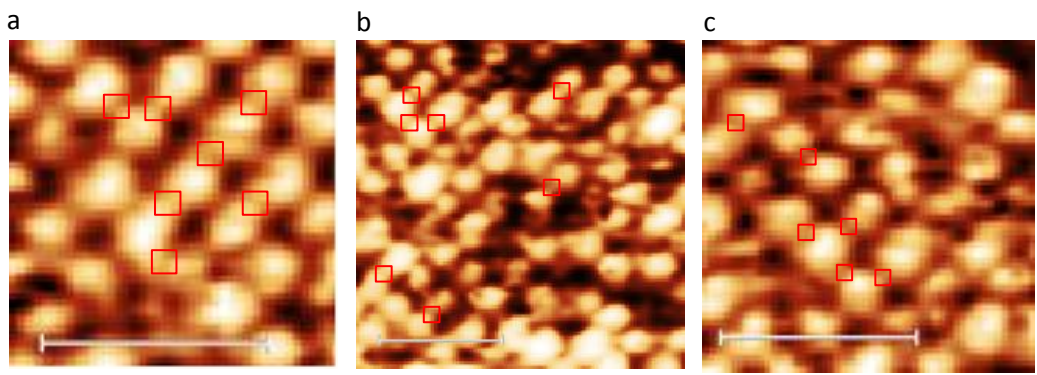

**Supplementary Figure 1:** Enlarged STM Images without cleaning (only contrast adjustment was performed), showing that the same features of the cleaned images are present in the raw images. The cleaning is a necessary step to enable accurate atom location determination, and to remove scars from tip changes during scanning. Scale bar in (a-c) is 2 nm. Some examples of the elongation along the diagonal directions (which would indicate locally higher electron density, and therefore greater likelihood of La atoms on these sites) are indicated as red squares in (a-c).

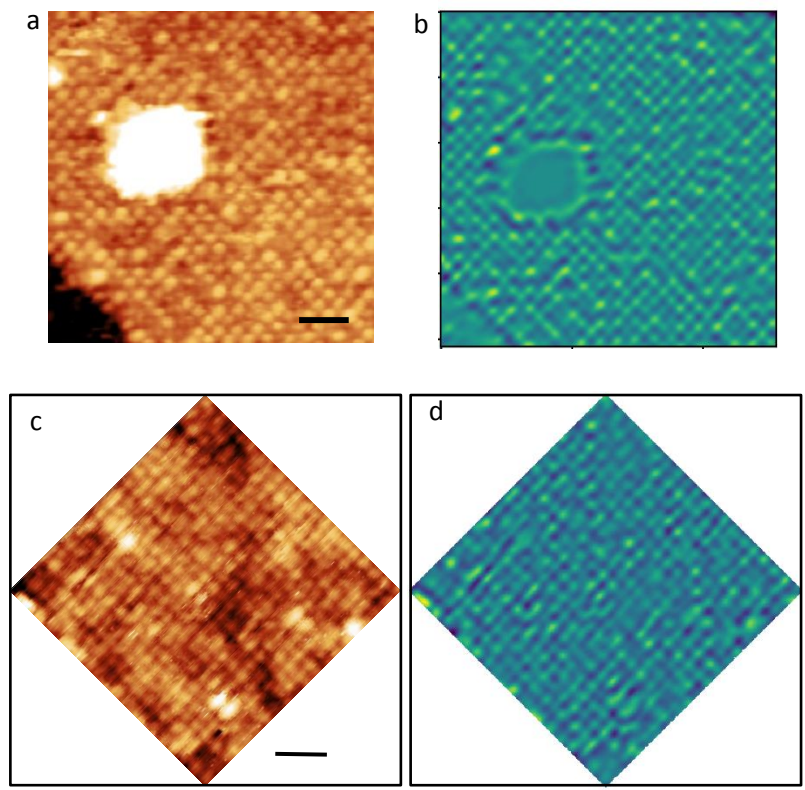

**Supplementary Figure 2:** PCA cleaned versions of two more STM images. Raw STM images in (a,c) and PCA-cleaned images in (b,d). Scale bar in (a,c), 2 nm.

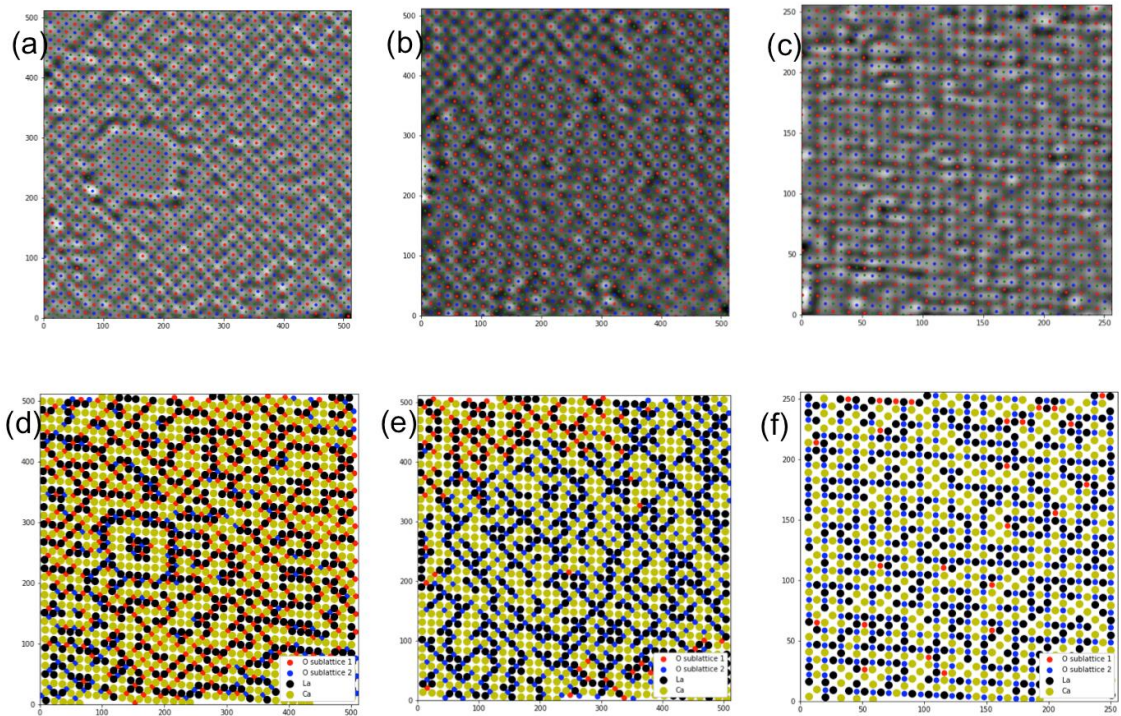

**Supplementary Figure 3:** Three images used for the statistical analysis of surface local configurations. **(a-c)** PCA-cleaned images with the highlighted ideal positions of oxygen and metal atom sites on identified lattices. Red and blue dots denote oxygen sites belonging to two sublattices, and green dots denote the positions of metal atoms. Note, the raw images for (a,c) are in Supplementary Figure 2, and the raw image for (b) is the one in the manuscript. **(d-f)** The reconstructions of surface structures based on the analysis of intensities at the ideal lattice sites. Red and blue dots denote O atoms, while black and yellow dots denote La and Ca, respectively.

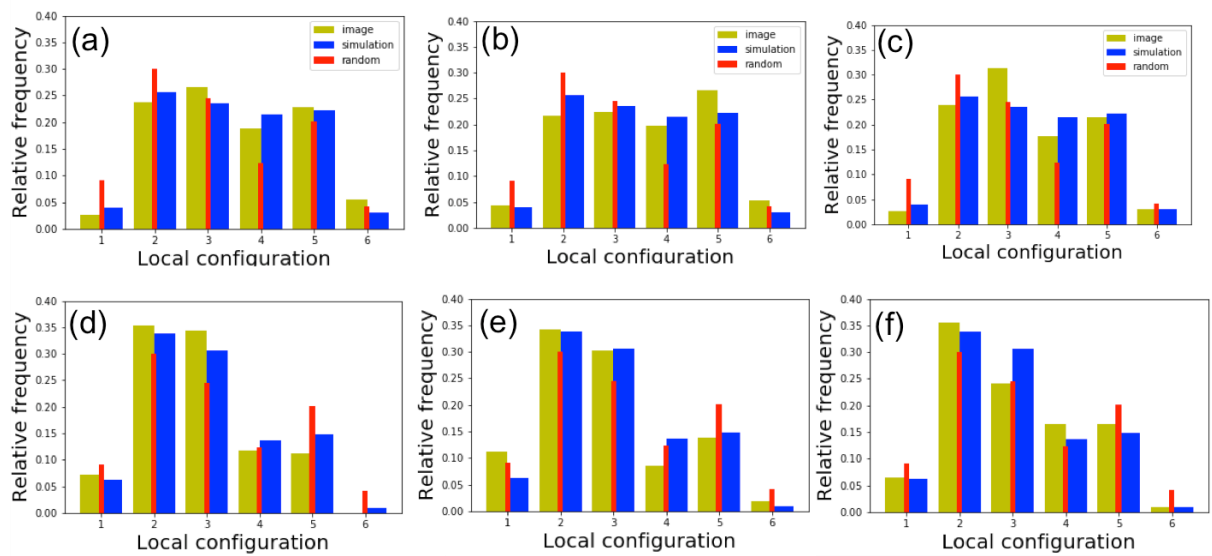

**Supplementary Figure 4: Structural statistics for LCMO thin film from experiments based on Figs. 3d-f (yellow) and simulations (blue).** Relative frequencies of configurations surrounding a surface oxygen **(a-c)**, and configurations without a surface oxygen **(d-f)**. Red bars denote relative frequencies of completely random configurations in a non-interacting system. Note that regions with clear defects (such as the circular region in Fig. 3a and 3d) were deliberately excluded from the analysis of local configuration statistics.

## Supplementary Note 1 - Statistical distance optimization

Statistical distances $s^2$ in Eq. (2) can be defined for any type of structural or thermodynamic property. In the present case of the statistics of local configurations $s_{SO}^2$ and $s_{SN}^2$, we can express the distance in terms of relative frequencies of the 6 configurations collected from the images, $\bar{p}_{i,SO}$ ($\bar{p}_{i,SN}$), and from the simulations, $\bar{q}_{i,SO}$ ($\bar{q}_{i,SN}$), as[61]

$$s_{SO}^2 = \arccos^2\left(\sum_{i=1}^{6}\sqrt{\bar{p}_{SO,i}}\sqrt{\bar{q}_{SO,i}}\right) \quad ; \quad s_{SN}^2 = \arccos^2\left(\sum_{i=1}^{6}\sqrt{\bar{p}_{SN,i}}\sqrt{\bar{q}_{SN,i}}\right). \tag{S1}$$

In case of the concentration profiles of $x$ = Ca/(La+Ca) fraction, we can assume that the instantaneous concentrations of Ca in individual layers follow the Gaussian distribution. Consequently, we can measure statistical distance between the experimental and simulated distributions as,[61]

$$s_{PR}^2 = \sum_{l=1}^{16}\arccos^2(C_l) \quad ; \quad C_l = \exp\left[-\frac{\left(x_l^{EXP} - x_l^{SIM}\right)}{8\left(\mathrm{D}x_l^{SIM}\right)^2}\right] \tag{S2}$$

Here the summation runs for the 16 layers of the thin film, $x_l^{EXP}$ and $x_l^{SIM}$ are the mean Ca concentration fractions in layer $l$ obtained from experiment and from simulations, and $\left(\mathrm{D}x_l^{SIM}\right)^2$ denotes variance of the Gaussian distributions, measuring the concentration fluctuations in each layer. Since these were not available from experiment, the value measured in simulations was used as an approximation.